# Cosmic Muon Explorer: A Portable Detector for Cosmic Muon Flux Measurements and Outreach

Yuvaraj Elangovan[1*], Shashwat Kakkad[2] and B. Satyanarayana[3]

[1]University of Pittsburgh, PA, United States
[2]Northwestern University, IL, United States
[3]Tata Institute of Fundamental Research, Mumbai, India
E-mail:
yue8@pitt.edu

**Abstract**

In this paper, we present development of a portable cosmic muon tracker tailored for both on-site measurements of cosmic muon flux as well as for outreach activities. The tracker comprises of two 70 mm × 70 mm plastic scintillators, wavelength shifting (WLS) fibers, and Hamamatsu made Silicon Photomultipliers (SiPM) S13360-2050VE. The detector uses plastic scintillator panels optically coupled to WLS fibers, which transmit scintillation light to the SiPMs. SiPM signals are routed to an electronics board equipped with op-amp amplifiers and a peak-hold circuit connected to an ESP32 microcontroller module. When muons traverse through both scintillators the light emitted is collected by the SiPMs, and thus generating signals proportional to the incident light intensity. These signals are then amplified and the pulse peak is held for 500 µs. A high speed discriminator is used to generate trigger logic signals. The peak analog voltage is digitized using the onboard ADC of the ESP32 when a coincident trigger occurs. The SiPMs are powered by a High Voltage bias supply module while an onboard BMP180 module measures temperature and pressure. For real-time event tagging, a GPS module is interfaced with the ESP-32. Housed within an acrylic box measuring 100 mm × 100 mm × 100 mm the detector can be powered using a 5V 1A USB power bank. An open source mobile application was used for real-time monitoring. This versatile and cost-effective portable detector facilitates cosmic muon research in various experimental settings. Its portability and low power requirements enable on-site measurements in environments such as tunnels, caves as well as high altitudes.

## 1. Introduction

Earth's atmosphere is continuously bombarded by high energy particles known as Cosmic rays originating from outer space. They are mostly composed of protons and heavier nuclei. When the cosmic rays interact with the atmospheric nuclei, they produce a cascade of secondary particles. This process is called as atmospheric shower or cosmic ray shower as shown in Figure 1. One of the results of these collisions are pions, which are very shortlived and therefore decay into muons and muon neutrinos. Muons are elementary particles similar to electrons but about 200 times heavier. They are minimum ionizing particles that interact weakly with matter compared to hadrons, which gives them strong penetrating capability. Their comparatively long lifetime [1] further allows many of them to travel large distances and reach the Earth's surface. This penetrating ability allows muons to be detected at sea level and even underground, making them valuable for various scientific applications.

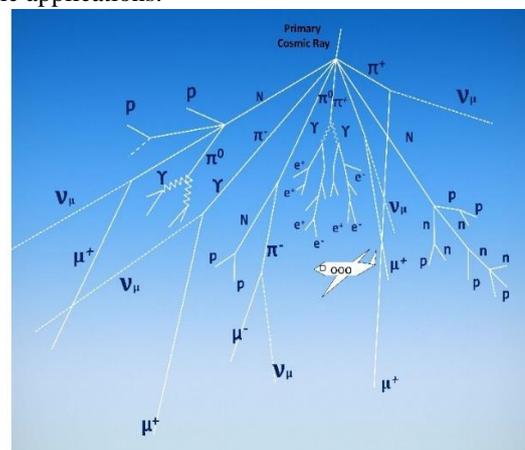

**Figure 1.** Cosmic Ray Shower in the earth's atmosphere.

Muon detection plays a crucial role in fields such as particle physics, atmospheric science, geophysics and environmental monitoring. In particle physics, muons are used to probe material properties and study fundamental interactions while in atmospheric science they provide insights into cosmic ray interactions and the composition of the upper atmosphere. In geophysics and environmental monitoring muon tomography is employed to image the internal structures of large objects such as volcanoes or pyramids. The development of the Cosmic Muon Explorer shown in Figure 2 is a portable detector, aimed at measuring cosmic muon flux on-site enabling studies in a wide range of environments and outreach [2]. This project was inspired by the Cosmic Watch Desktop Muon Detector a compact and small scale detector unit designed for muon detection [3].

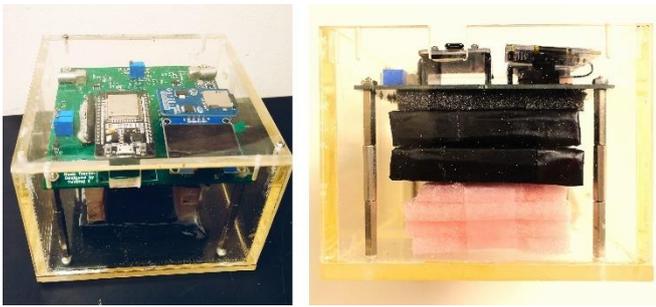

**Figure 2.** Cosmic Muon Explorer Detector module.

Plastic Scintillators can be used in coincidence to study Muon flux variations [4]. By utilizing plastic scintillators coupled with Silicon Photomultipliers (SiPMs) and an efficient readout system, the detector can accurately capture and analyze muon events. Its portability allows for deployment of the detector module in diverse settings from high-altitude locations to subterranean environments, offering valuable data on muon flux and its variations. This data is not only essential for scientific research but also serves practical applications such as environmental monitoring and structural imaging. The Cosmic Muon Explorer thus represents a robust tool for hand held muon detection and cosmic ray studies. The muon tracker consists of plastic scintillators, wavelength shifting (WLS) fibers, Silicon Photomultipliers (SiPMs) and simplified data acquisition system to accurately measure and record muon flux. The core of the muon tracker consists of two plastic scintillators coupled with WLS fibers, which guides scintillation light to SiPMs (Hamamatsu S13360-2050VE) [3]. These SiPMs are connected to an electronics circuit board (PCB) that includes operational amplifiers, peak hold circuit and a discriminator. The muons passing through the scintillators generate light pulses that are detected by the SiPMs. The electrical signals from the SiPMs are amplified and peak held for a duration of 500 µs and digitized using an ADC inside the ESP32 microcontroller module. The ESP32 records these signals and flags as a muon event when the trigger signals from both discriminators turn active. A BMP180 sensor module was used to record temperature and pressure along with muon events. Also an MPU6050 based accelerometer is used to keep track of relative angle of the coincidence measurement. The entire system is powered by a 5V 1A USB power bank and can be monitored in real-time via a mobile application.

The Cosmic Muon Explorer prototype was initially developed at the Tata Institute of Fundamental Research, India and the major system development and integration were subsequently carried out at the University of Pittsburgh, United States. The present system is not intended as a classroom construction project, but rather as a compact particle detector designed primarily for live demonstrations, undergraduate laboratory use, and guided outreach activities. This pilot effort focuses on establishing a reliable miniature detector platform that can operate in diverse environments such as university laboratories, schools, and public demonstrations. Based on the experience gained from this prototype, future work is directed toward developing ready-made detector kits that will allow students to assemble the system more easily and perform particle detection studies in versatile educational settings. This paper describes design and fabrication of the detectors, electronics, trigger and data acquisition systems, performance of the stack as well as its potential applications.

## 2. Geometry of the Detector Module

The detector uses two plastic scintillators as shown in Figure 3 (Left) each measuring 70 mm × 70 mm × 10 mm, stacked on top of each other. This configuration allows for the detection of muons simultaneously passing through both scintillators, and thus minimising the chance coincidence. Size of the scintillators is chosen to ensure portability while still providing sufficient muon detection rate and efficiency.

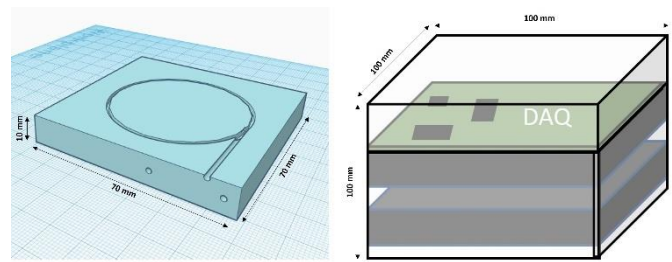

**Figure 3.** Detector Dimensions and Enclousure.

The electronics board is placed on top of the two detectors using metal studs. The entire setup is enclosed in an acrylic box of dimensions of 100 mm × 100 mm × 100 mm as shown in Figure 3 (Right). Slots were made into the box to support USB cable connection for powering and readout.

## 3. Detector Fabrication

### 3.1 Scintillator Machining

Small scintillator slabs were cut from a larger scintillator [4] block using a vertical band saw. A 50 mm diameter and 2 mm deep groove was milled into the surface of the scintillators as shown in Figure 4 (Left) to hold the WLS fiber. The edges of the scintillators were left rough due to the saw blades which negatively impacts light transmission. To address this, all the edges of the scintillator were polished using dry sandpaper or lapping sheets of various grits between 200 and 2500 as shown in Figure 4 (Right).

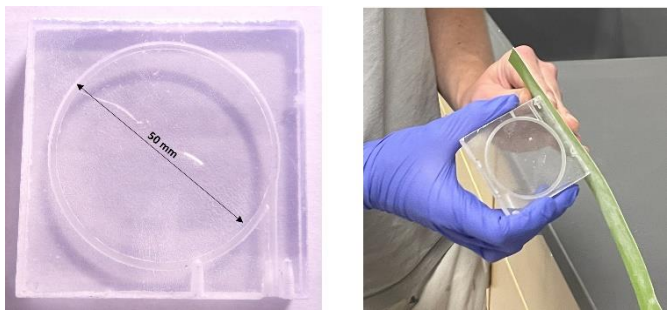

**Figure 4.** Scintillator machining and polishing.

During sanding and polishing it is important to be cautious of heat buildup. Excessive heat can warp or damage the scintillator, thus reducing its performance. Even pressure was applied to avoid making uneven surfaces. It is also recommended to wear gloves to prevent body oils contaminating the polished surface which can degrade its optical clarity. An N95 mask must be worn to protect against dust particles generated during sanding. Properly polished scintillators will ensure maximum light output and optimal detector performance.

### 3.2 WLS Fiber Polishing

Kuraray Wave Length Shifting (WLS) Y-11 (250) fiber [5] were cut to the required length using razor blade. However, since the cuts were made manually, the surfaces of the fiber ends are not smooth as shown in Figure 5. To address this, both ends of the fiber were required to be polished. This was achieved using a range of lapping sheets from grit size of 600 to 2500 as shown in Figure 5. The fibers are held close to the polishing end to minimize stress, while polishing is done in a figure-eight pattern to reduce strain on the fiber tip.

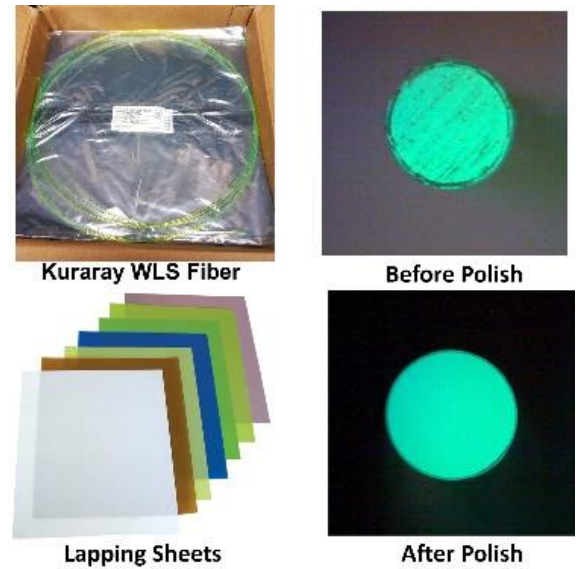

**Figure 5.** WLS FIber polishing using grit sheets.

At each stage, the polishing quality was monitored by observing the polished surface under a lab microscope. A well polished fiber end will have a shiny glass-like appearance.

### 3.3 SiPM Carrier Module and SiPM Fiber Guide (SFG) Block

The SiPM carrier module as shown in Figure 6 (Left) is a compact printed circuit board (PCB) of dimensions of 30 mm × 10 mm. This PCB hosts the SiPM on one side while the other side contains the readout connector and bias filter. A 3-pin header representing Signal, Ground and Bias is used for readout.

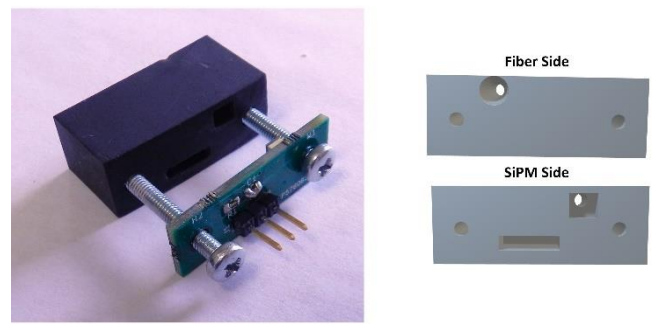

**Figure 6.** SiPM Carrier Board and Fiber Guide.

The SiPM fiber guide (SFG) shown in Figure 6 (Right) was designed as a 3D model and manufactured using a 3D printer supporting precise alignment between the SiPM and the WLS fiber. The guide holes facilitate secure and efficient coupling optimizing light collection from the scintillator to the SiPM. Dedicated holes and slots are incorporated into the SFG to ensure a firm connection between the scintillator, SFG and the SiPM carrier module.

*3.4 Detector Packing*

In Figure 7, we show the step-by-step assembly and packaging procedure of the scintillators and their interface components. After machining the scintillators, the polished fiber is placed into the groove leaving 10 mm of the fiber exposed. The SFG is then positioned to align the fiber close to the SiPM. The SiPM carrier module along with the SFG is attached to the scintillator using two M3 screws supporting precise alignment between the SiPM and the WLS fiber. The guide holes facilitate secure and efficient coupling optimizing light collection from the scintillator to the SiPM. Dedicated holes and slots are incorporated into the SFG to ensure a firm connection between the scintillator, SFG and the SiPM carrier module.

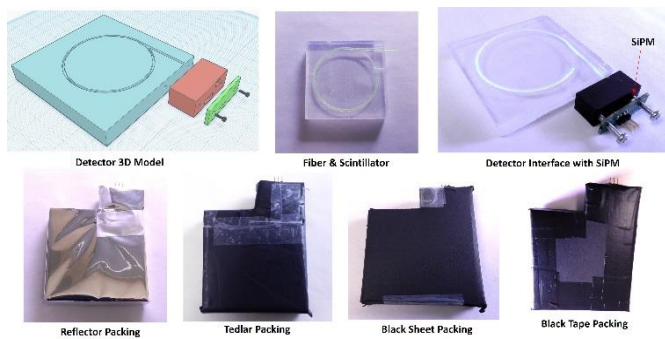

**Figure 7.** Step by Step Procedure of Detector Packing.

After coupling the WLS fiber to the scintillator, each scintillator plate is first wrapped with a reflective layer to improve internal light collection. This reflective covering helps redirect scintillation photons toward the fiber and SiPM, thereby enhancing the signal yield. The assembly is then enclosed in a black light-tight wrapping, typically using Tedlar film, black sheet and black tapes to prevent ambient light from reaching the detector and to ensure stable detector operation.

## 4. SiPM Characterization

The detector uses the Hamamatsu S13360-2050VE silicon photomultiplier operates at a typical bias voltage of approximately 54 V (depending on temperature and overvoltage setting). The sensor provides a peak photon detection efficiency of about 40% in the blue-green wavelength region relevant for plastic scintillators, with a typical dark count rate of order 300 kHz at room temperature.

The SiPMs were characterized using an LED pulser inside a black box to mimic scintillation light conditions [6]. A narrow 10 ns wide pulse was used to trigger both the LED and the oscilloscope. The resulting signal from the SiPM was amplified and fed into the oscilloscope. For each LED trigger pulse the amplified SiPM response waveform was recorded by the oscilloscope. The charge of each pulse was calculated by measuring the area under the curve of the waveform. The charge distribution from the SiPMs exhibited distinct photoelectron peaks as shown in Figure 8, each corresponding to the number of detected photons. These distinct peaks and the spacing between them are essential for accurately calibrating the SiPMs. This process was repeated at different bias voltages ranging from 53V to 55V and it was determined that the SiPMs showed the highest efficiency at 54V.

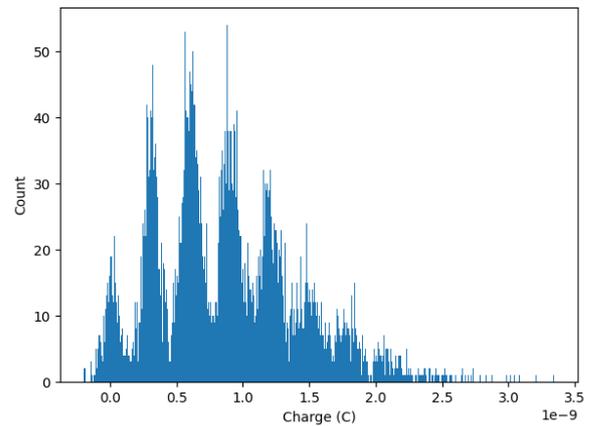

**Figure 8.** Photo electron peak distribution of an SiPM.

## 5. Trigger and Data Acquisition

Two scintillators are placed one on top of the other with their signal coaxial cables connected to the motherboard. The motherboard as shown in Figure 9 was designed using an ESP32 microcontroller along with an amplifier, peak holder, discriminator and SiPM bias circuits as well as temperature and pressure sensors. The motherboard was designed as a 4-layer PCB where the top and bottom layers were used for signal routing, and the second and third layers serve as ground and power planes respectively.

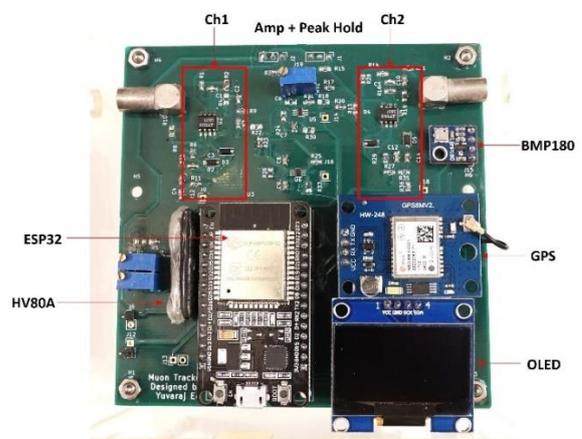

**Figure 9.** DAQ Mother Board.

Figure 10 shows a block diagram of the readout signal scheme of the module. Signals from the SiPMs are first amplified using a voltage amplifier with a gain of 30. For debugging, these amplified signals can be connected to an oscilloscope through two LEMO connectors. A peak-holder operational amplifier stage follows the amplification stage to capture and retain the maximum amplitude of the SiPM pulse. The circuit is implemented using an op-amp peak detector configuration with two diodes and a hold capacitor. The hold time is set by the RC time constant and is adjusted to approximately 500 µs, which is sufficient for reliable sampling by the ADC. This approach allows the fast scintillation pulses to be digitized accurately while maintaining acceptable recovery time for subsequent events as shown in Figure 11 (Left). The amplified signal is also converted to a digital logic trigger signal shown in Figure 11 (Right) using a discriminator with a reference voltage of 100 mV which was set by a potentiometer.

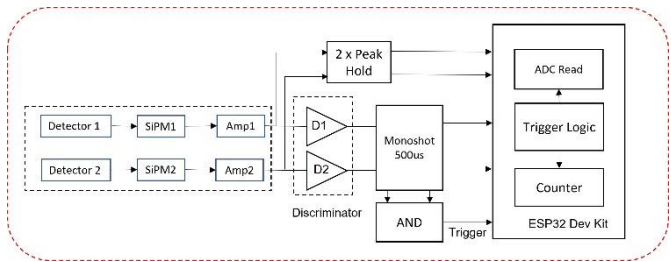

**Figure 10.** Block diagram of the readout signal scheme

The trigger signals from both channels are stretched to 1 µs using a pulse stretcher and then fed into a logic AND gate to produce a coincidence trigger signal. The data acquisition (DAQ) system is based on the ESP32 microcontroller [7] which processes signals and also logs data. When a coincidence trigger is detected, the ESP32 digitizes the peak held signals and increments the coincidence count. Also the individual trigger signals from each channel are counted and stored in memory. An alternative, simpler method of operation involves continuously digitizing the peak signals and comparing them to a fixed threshold. If the peak values from both channels exceed certain threshold, then the event is considered as coincidence.

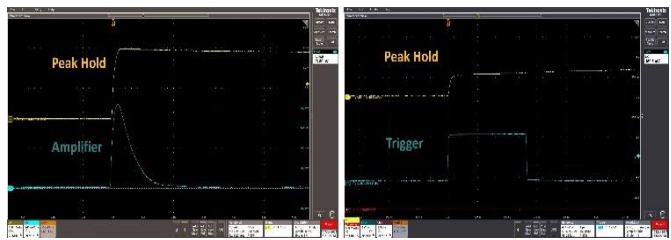

**Figure 11.** Detector signals measured on oscilloscope.

The temperature and pressure data are periodically recorded by the BMP180 sensor module and an onboard MPU6050 accelerometer module measures the detector's angle during measurements. All measured values are displayed on onboard OLED display. A GPS interface is optionally available for recording real-time coordinates during measurements. The SiPMs require bias voltage to function in Geiger mode which is supplied using HV80A DC-DC converter. This converter ensures stable and low-noise voltage whose output voltage is adjusted via a trim pot. Additionally, the HV80A module includes a fail-safe mechanism that turns off the power in case of over-current detection.

## 6. Run Control and Data Logging

The ESP-32 module operates using a C code developed using Arduino IDE. The code includes a programmable timer that collects data over a specified interval - typically 10 seconds.

**Table 1.** Event Data Format.

| Datatype | Description |
|---|---|
| Peak1 | Peak value of Scintillator1 |
| Peak2 | Peak value of Scintillator2 |
| Count1 | Scintillator1 counter value |
| Count2 | Scintillator2 counter value |
| Coincidence | Coincidence trigger counts |
| Temperature | Temperature on-site |
| Pressure | Pressure in mbar |
| Altitude | Altitude in meters |
| Angle | The Y axis angle in deg |

At the end of each timer interval a data frame is created containing all the acquired information as shown in Table 1. This data frame is then transmitted to the back-end via wireless Bluetooth [8] or wired serial communication links.

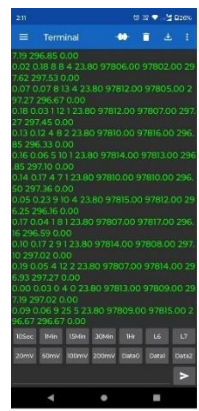 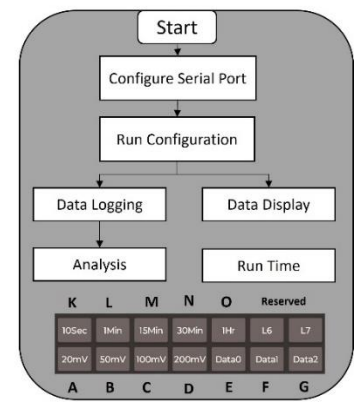

**Figure 12.** Serial Bluetooth terminal and block diagram of the Python based back-end software.

After transmission, all counters are reset to zero and the system returns to the acquisition mode. This process repeats based on the user-programmed timer interval. Data from the Cosmic Muon Explorer can be collected and monitored using various devices such as laptops, desktops or mobile applications. The mobile app enables real-time monitoring of muon detection events displaying information like peak values, temperature, pressure and coincidence counts. This method of data logging and control system makes the detector suitable for both fixed-site experiments and portable field studies.

**Figure 13.** Python based back-end software executed from a terminal

For standard data collection, a Python-based backend software as shown in Figure 12 was developed to communicate with the ESP-32 over a serial interface at baud rate of 115200. At the start of a run, the user is prompted to input operating parameters such as the COM port number, signal threshold, monitoring period and data type as shown in Figure 13 . The monitoring period can be set between 10 seconds and 30 minutes. Additionally, the data type field allows users to select the output format with three options: a complete data set, a simplified format with only the coincidence count or a format displaying essential parameters such as counts, sensor values as well as angle of the detector orientation. These parameters can be configured in real-time by entering the respective alphabet commands as illustrated in Figure 12.

## 7. Test Results and Measurements

### 7.1 Coincidence Measurement

The Cosmic Muon Explorer was subjected to a series of tests to validate its performance. A dedicated software based on Python and Matplotlib was developed to analyse the data for plotting. The test results from the Cosmic Muon Explorer collected over a period of approximately 36 hours are shown in Figure 14 and Figure 15.

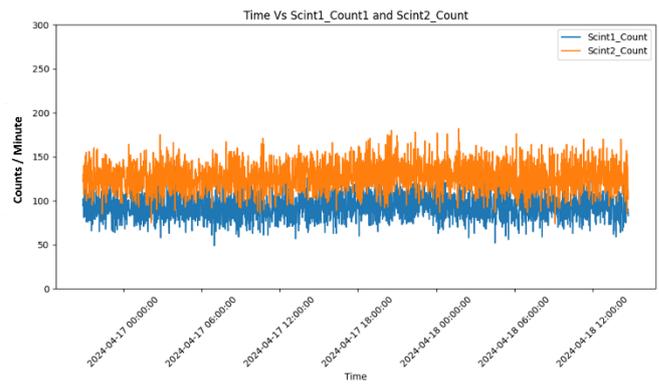

**Figure 14.** Individual scintillator count rates (counts per minute) for Scint1 and Scint2 as a function of time. Each point represents data averaged over 1 minute

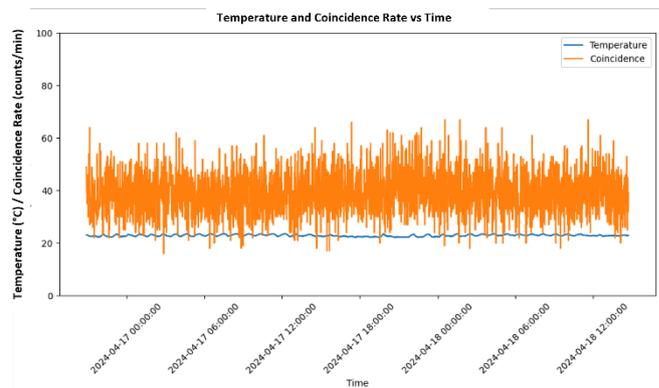

**Figure 15.** Temperature (°C) and coincidence rate (counts per minute) as a function of time. Each point represents data averaged over 1 minute.

The Figure 14 illustrates the scintillation count rates for two scintillators (Scint1 and Scint2) over time. The vertical scale represents counts per minute, obtained by averaging the monitoring data over 1-minute intervals. The observed single rates are consistent with expectations based on the known vertical cosmic muon flux of approximately 150–200 m$^{-2}$ s$^{-1}$ at sea level. For a scintillator area of about 7 × 7 cm² the expected muon rate is of order 40–60 counts per minute. The elevated individual scintillator counts are due to high gain amplification and approximate discriminator threshold to eliminate electronic noise and accept valid signals.

Figure 15 shows the variation of temperature and coincidence count rate over time during the monitoring period. No significant dependence of the coincidence rate on temperature is observed over the measured range. As expected, the coincidence measurement is insensitive to small temperature variations, although SiPM single channel rates can exhibit stronger temperature dependence.

### 7.2 Testing with radioactive source

The ⁶⁰Co source was placed between two detectors while acquiring the data. The ⁶⁰Co source emits two nearly simultaneous gamma rays (1.17 MeV and 1.33 MeV), which can produce correlated interactions in the two scintillators. The purpose of this measurement was to verify the detector response to such correlated gamma events and to evaluate the stability of the coincidence counting over extended running periods. From the observed average count rate (in counts per minute), the expected relative statistical fluctuation is of order $1/\sqrt{N}$, which is consistent with the spread observed in the time series. Since the detector bias, geometry, and environmental conditions remained stable during the few-hour acquisition period, the coincidence rate is expected to remain constant within statistical uncertainties, as observed.

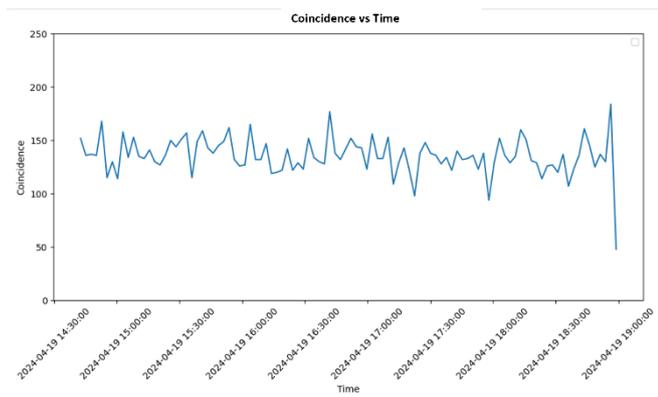

**Figure 16.** Measured Coincidence count with Co-60 Source over time.

### 7.3 Angular distribution of muons

To measure the angular distribution of muons, the detectors were spaced 500 mm apart [9] and rotated from vertical to horizontal orientation using a table top custom made stand as shown in Figure 17 (Left). The Accelerometer MPU6050 was placed in between the two detectors to record accurate angle during muon flux measurement.

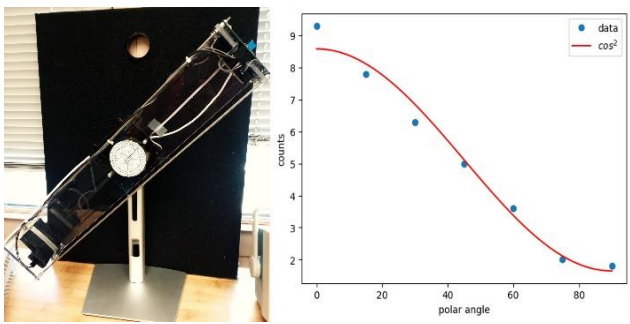

**Figure 17.** Experimental setup with adjustable polar angle, and measured muon count rate as a function of polar angle with a Cosine Squared fit.

Overnight measurements were taken at each position to gather sufficient data. The angular flux distribution was analyzed and fitted to a Cosine Squared function confirming the expected distribution pattern of cosmic muons shown in Figure 17 (Right).

### 7.4 Flux Measurement inside railway tunnel

The Cosmic Muon Explorer was also tested in a real-world scenario by taking measurements as it was carried on a train between Mumbai and Goa in India. The detector recorded muon flux variations as it passed through a series of tunnels as shown in Figure 18, thus demonstrating its capability to measure cosmic muon flux on field. It was observed that due to the detector's small form factor and the movement of the train through tunnels the coincidence count drops to zero while passing through most of the tunnels. To obtain more meaningful data, it is essential to keep the detector stationary at various points within the tunnel. This approach has been proposed for future measurements.

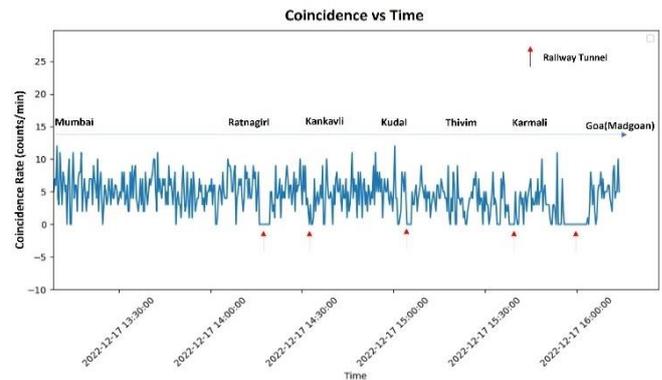

**Figure 18.** Coincidence rate (counts per minute) as a function of time during the tunnel measurement. Red arrows indicate the time intervals when the train was inside the tunnel.

### 8. Conclusion

Currently one unit of this detector was developed and studied for real field challenges. One notable issue with this prototype is rise of temperature due to enclosed acrylic box. This issue will be addressed in the next version. Cosmic Muon Explorer's compact size, low power requirements and ease of use makes it an ideal tool for on-site muon flux measurements in diverse environmental conditions. The successful tests in both controlled and real-world scenarios highlight its potential for widespread scientific, educational and outreach applications.

## Acknowledgements

We sincerely thank Mandar Saraf, Ravindra Shinde, Raj Shah, Darshana Gonji, Santosh Chavan, and Vishal Asgolkar for their continuous support during the prototype development. We also express our gratitude to former INO Director V. M. Datar for his encouragement and guidance. We further thank Danko Istvan and David Emala for their support during the major development phase. Finally, we acknowledge the Department of Physics and Astronomy and the Electronics Shop at the University of Pittsburgh for their support in the development of the detector used for angular distribution measurements.

## References

[1] Peter Dunne and David Costich and Sean O'Sullivan, Measurement of the mean lifetime of cosmic ray muons in the A-level laboratory, *Physics Education,* **33** (1998) 5

[2] Tejeda Muñoz et al., Cosmic Piano: a modular scintillator-based muon detector for scientific outreach, *Physics Education,* **60** (2025) 3

[3] The CosmicWatch Desktop Muon Detector: a self-contained, pocket sized particle detector, *Journal of Instrumentation,* **13** (2018)

[4] Muon Flux Variations Measured by Low-Cost Portable Cosmic Ray Detectors and Their Correlation With Space Weather Activity, *Journal of Geophysical Research,* **128** (2023) 12

[5] SiPM S13360-2050VE, https://www.hamamatsu.com/us/en/product/optical sensors/mppc

[6] Plastic Scintillators, https://eljentechnology.com/products/plastic-scintillators

[7] Wave Length Shifting Fiber, KURARAY, JAPAN, https://www.kuraray.com/products/psf

[8] Jangra, Mamta and others, Characterization of Hamamatsu SiPM for Cosmic Muon Veto Detector at~IICHEP, *Springer Proc. Phys,* **277** (2022) 815-819

[9] Espressif ESP-32 Development Kit, https://www.espressif.com/en/support/documents/technical-documents

[10] Serial Bluetooth Terminal, https://github.com/kai-morich/SimpleBluetoothTerminal?tab=readme-ov-file

[11] M. Bahmanabadi, A method for determining the angular distribution of atmospheric muons using a cosmic ray telescope, *Nuclear Instruments and Methods in Physics Research Section A: Accelerators, Spectrometers, Detectors and Associated Equipment,* **916** (2019) 1-7